\documentclass[reprint,prb,twocolumn,longbibliography,noeprint,superscriptaddress,floatfix]{revtex4-2}
\usepackage{float}
\usepackage{physics}
\usepackage{graphicx}
\usepackage{xcolor}
\usepackage{amsmath,,amsfonts,amssymb}
\usepackage{textcomp}
\usepackage{bbm}
\usepackage{dsfont}
\usepackage{microtype}
\usepackage{mathtools}
\usepackage{multirow}

\usepackage[pdftex,bookmarks=true,bookmarksopen,bookmarksnumbered,
                colorlinks,
                linkcolor=blue,
                citecolor=blue,
                colorlinks = true,
                urlcolor  = blue,
                anchorcolor = blue
                ]{hyperref}
\usepackage{cleveref}

\renewcommand{\figurename}{FIG.}
\makeatletter
\renewcommand*{\fnum@figure}{{\normalfont \figurename~\thefigure}}
\renewcommand*{\@caption@fignum@sep}{ $~$}

\renewcommand{\tablename}{Table}
\makeatletter
\renewcommand*{\fnum@table}{{\normalfont \tablename~\thetable}}

\usepackage{float}
\usepackage{newfloat}
\usepackage{standalone}
\usepackage{booktabs}
\usepackage[geometry]{ifsym}

\crefname{figure}{Fig.}{Figs.}
\crefname{table}{Table}{}
\crefname{section}{Sec.}{Secs.}
\crefname{equation}{Eq.}{Eqs.}

\graphicspath{{paper_figures/}}

\newcommand{\si}{\texorpdfstring{\textsuperscript{29}Si}{si29}}
\newcommand{\siv}{\texorpdfstring{\textsuperscript{29}SiV\textsuperscript{-}}{si29}}

\newcommand{\vect}[1]{$\boldsymbol{\rm{#1}}$}

\usepackage{lineno}
\DeclarePairedDelimiter\ceil{\lceil}{\rceil}

\begin{document}

\title{Indirect Control of the \texorpdfstring{\textsuperscript{29}SiV\textsuperscript{-}}{29SiV} Nuclear Spin in Diamond}

\author{Hyma H. Vallabhapurapu}
\email[]{h.vallabhapurapu@unsw.edu.au}
\affiliation{
 School of Electrical Engineering and Telecommunications,
 The University of New South Wales, Sydney, NSW 2052, Australia
}

\author{Chris Adambukulam}
\affiliation{
 School of Electrical Engineering and Telecommunications,
 The University of New South Wales, Sydney, NSW 2052, Australia
}

\author{Andre Saraiva}
\affiliation{
 School of Electrical Engineering and Telecommunications,
 The University of New South Wales, Sydney, NSW 2052, Australia
}

\author{Arne Laucht}
\email[]{a.laucht@unsw.edu.au}
\affiliation{
 School of Electrical Engineering and Telecommunications,
 The University of New South Wales, Sydney, NSW 2052, Australia
}

\begin{abstract}
 Coherent control and optical readout of the electron spin of the \siv{} center in diamond has been demonstrated in literature, with exciting prospects for implementations as memory nodes and spin qubits. Nuclear spins may be even better suited for many applications in quantum information processing due to their long coherence times. Control of the \siv{} nuclear spin using conventional NMR techniques is feasible, albeit at slow kilohertz rates due to the nuclear spin's low gyromagnetic ratio. In this work we theoretically demonstrate how indirect control using the electron spin-orbit effect can be employed for high-speed, megahertz control of the \si{} nuclear spin. We discuss the impact of the nuclear spin precession frequency on gate times and the exciting possibility of all optical nuclear spin control.
\end{abstract}

\maketitle

\section{\label{sec:introduction} Introduction}

\begin{figure*}[htbp]
\includegraphics[keepaspectratio]{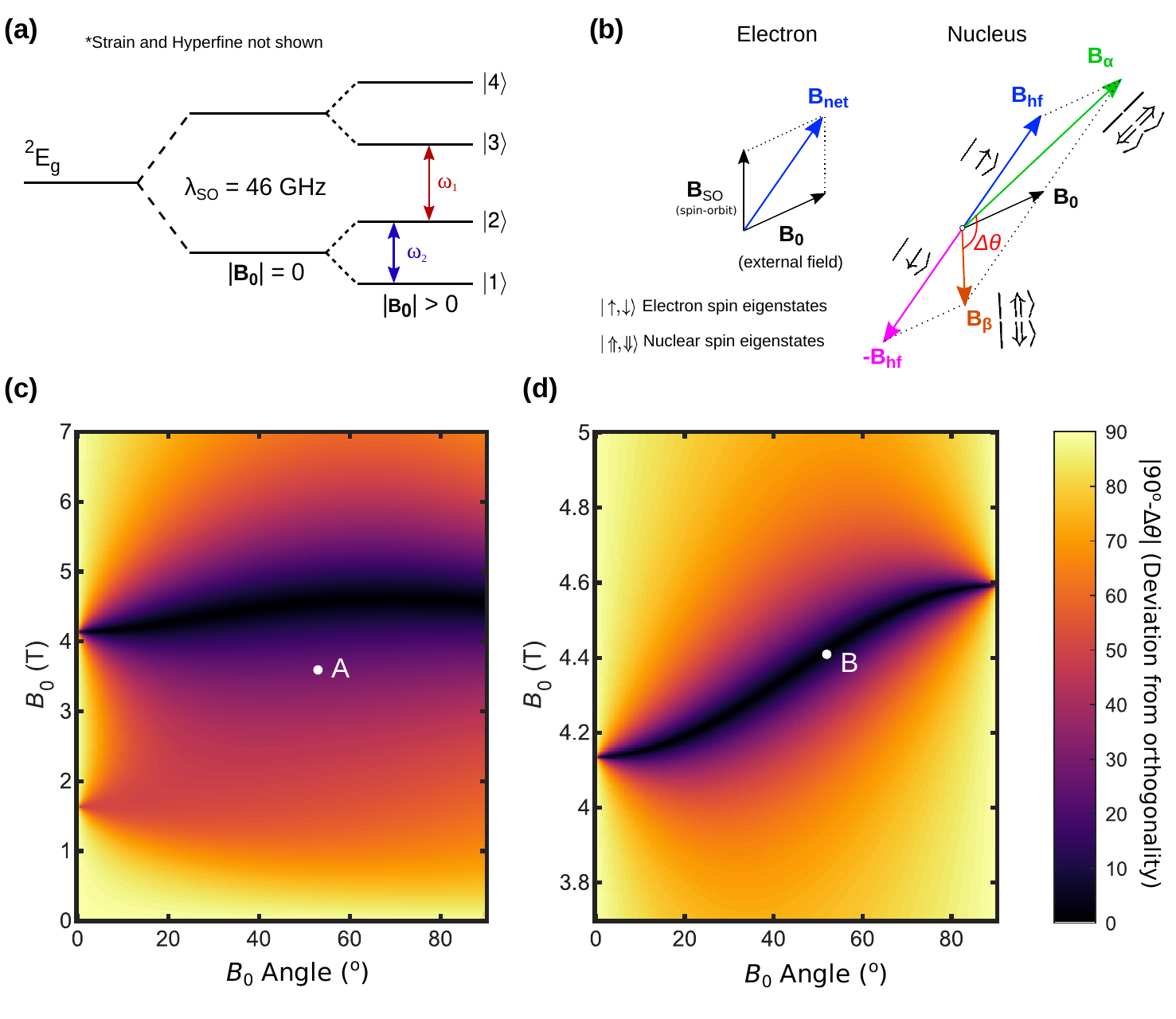}
\caption{Nuclear spin quantization field.
(a) Ground state energy levels showing Zeeman-split electron spin states. Relevant electron spin transitions for the zero-strain and moderately strained cases are shown as red and blue arrows, respectively. 
(b) The primary \vect{B_{\rm \alpha}} and secondary \vect{B_{\rm \beta}} quantization fields represent the net fields (\vect{B_{\rm hf}} + \vect{B_0}) experienced by the nuclear spin as a result of the electron spin orientations. Here, $\Delta\theta$ quantifies the change in the net quantization field direction of the nuclear spin as a result of electron spin $\pi$-rotations.
(c) Plot of $|90^{\circ{}}-\Delta\theta|$ as a function of the \vect{B_0} vector for the zero-strain case,  illustrating the deviation from orthogonal nuclear spin eigenstate orientations. The dark region represents the ideal operating parameters for the \vect{B_0} field so as to maximize $\Delta\theta$. 
(d) Same as (c) but for the moderately strained case. The labels `A' and `B' indicate the \vect{B_0} set-point parameters used for the simulations described in \cref{table_gate_times} in \cref{qubit_gates}.}
\label{b0_optimimum_plot}
\end{figure*}

The state of the art in diamond-based spin qubit implementations includes low temperature nanophotonics and coherent control of negatively charged group IV vacancies, as a result of their favourable optical and coherence properties at low temperatures ($\le$4~K) \cite{Bradac2019,Siyushev2017b,debroux2021quantum,Aghaeimeibodi_2021}. In particular, much progress has been made with the negatively charged silicon vacancy (SiV\textsuperscript{-}) color center in diamond, which has emerged as a promising candidate for the implementation of a spin qubit \cite{Rogers2014a,Sukachev2017,Becker2017b, Pingault2017,Nguyen,Bhaskar2020a}. 

Both electron and nuclear spins have long been proposed for applications as qubits for quantum information processing \cite{Loss1998,Kane1998a,Wrachtrup2001,Wrachtrup2006}. Nuclear spins are particularly suited for such applications due to their excellent isolation from the environment, resulting in long coherence times. Typical approaches to access the nuclear spins in diamond color centers involve finding a vacancy that is either hyperfine coupled to a nearby randomly occurring \textsuperscript{13}C atom \cite{Jelezko2004a,Smeltzer2011,Metsch2019,abobeih2019atomic} or to the host nuclear spin intrinsic to the vacancy center \cite{Scharfenberger2014,Rogers2014a}. A relevant example is the \siv{} center consisting of a \si{} isotope with a nuclear spin $I=1/2$ \cite{Pingault2017,Rogers2014a}, which is the focus of this work. Such intrinsic nuclear spins are preferred for scalability \cite{Fuchs2011} and take advantage of available precision defect implantation techniques in diamond \cite{Pezzagna2010,Schukraft2016}.

Fast, coherent control of the \si{} nuclear spin combined with the favourable optical readout properties of the SiV\textsuperscript{-} centers are envisioned to present robust implementations of spin qubit platforms for information processing. Multiple methods exist for coherent nuclear control, including conventional nuclear magnetic resonance (NMR) and recently nuclear electrical resonance (NER) \cite{Asaad2020}. However, these control methods are usually slow, and fast nuclear spin control of the \siv{} system presents a practical problem worth addressing. Firstly, electrical control is only possible for nuclei with spin $I>1/2$, eliminating the possibility for use with \siv{}. Secondly, the low gyromagnetic ratio of the host nuclear spin is roughly three orders of magnitude smaller than that of the electron, restricting the achievable nuclear Rabi frequencies using direct RF control to the kilohertz range. This is especially true for low temperature operation, where the usable microwave power is limited due to the maximally permissible thermal load of the cryostat. In contrast, electron spins can be driven at megahertz Rabi frequencies \cite{Fuchs2009,vallabhapurapu2021fast,Maity2020,Becker2018b}, and approaches exist to exploit the relative ease of high speed electron spin control to actuate on the nuclear spin towards a desired state via the hyperfine interaction, in a technique called \textit{indirect control} (IC) \cite{Khaneja2007,Zhang2019,Hegde2020}.

In the past, indirect control has relied on the anisotropic hyperfine interaction as a result of electron-nuclear dipolar interaction with nearby \textsuperscript{13}C nuclei \cite{Hodges2008,Nguyen2,maity2,takou2022multipartite}. However, the hyperfine coupling in the \siv{} defect is known to be dominated by the isotropic contact hyperfine term \cite{Pingault2017}; the anisotropic contribution to the hyperfine coupling is $\sim$2.7 MHz, which is relatively minor when compared to the isotropic contribution of $\sim$75.3 MHz \cite{Nizovtsev2020,Goss2007}. Fortunately, the \siv{} center possesses a significant spin-orbit coupling ($\sim$46 GHz \cite{Rogers2019}), due to which the electron-nuclear spin system exhibits similar anisotropic behaviour under an applied magnetic field \vect{B_0} -- making indirect control a possibility. 

In this work, we show via simulation of the spin Hamiltonian that fast control of the \siv{} host nuclear spin is possible using the IC technique. We first explore the parameters of the \siv{} Hamiltonian that enable indirect control of the nuclear spin in \Cref{sec_hamiltonian}. Since strain has a particularly significant impact on how the SiV qubit is implemented \cite{Meesala2018,Rogers2014a}, we explore both the unstrained and strained cases. In \Cref{sec_indirect} we use simulations to explicitly demonstrate several single qubit IC gates. Lastly, in \Cref{discussion} we discuss the effect of the \vect{B_0} field intensity and angle on the gate durations, along with possible applications of IC implemented using all-optical spin control methods.

\begin{figure*}[htb]
\includegraphics[keepaspectratio]{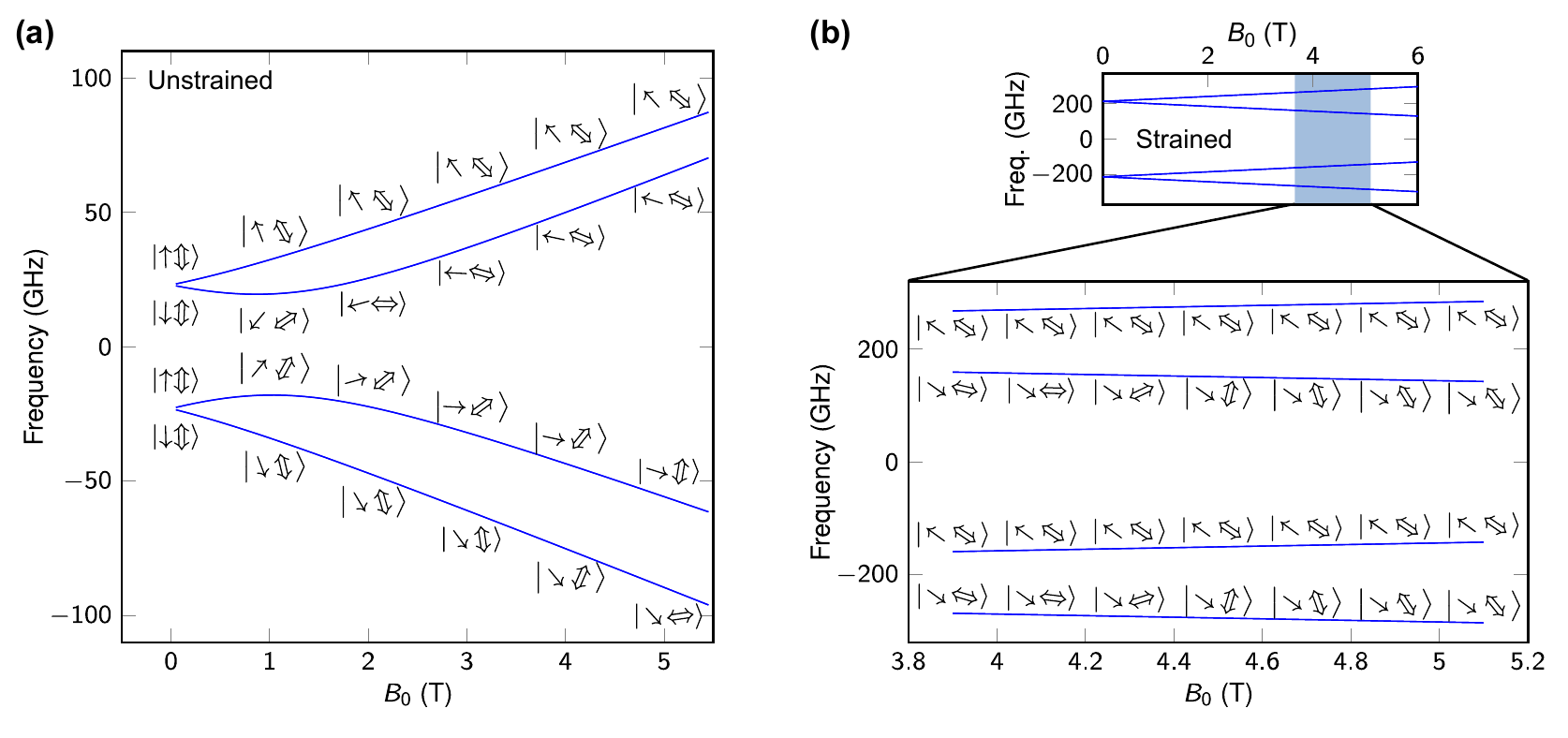}
\caption{Eigenstate orientations for the electron $\ket{\uparrow,\downarrow}$ and nuclear $\ket{\Updownarrow}$ spins. The $\ket{\Updownarrow}$ symbol indicates the presence of both the $\ket{\Uparrow}$ and $\ket{\Downarrow}$ nuclear spin states, as they are always oriented anti-parallel.
(a) Plot of the energy levels and spin eigenstate orientations for the zero-strain case, as a function of the \vect{B_0} field applied at an angle of $54.7^{\circ{}}$ away from the SiV axis. 
(b) Same as (a) but for the 150 GHz strain case. Note that the hyperfine splitting is not visible due to the scale of the vertical axes. }
\label{quant_axes}
\end{figure*}

\begin{figure}[hbp]
\includegraphics[keepaspectratio]{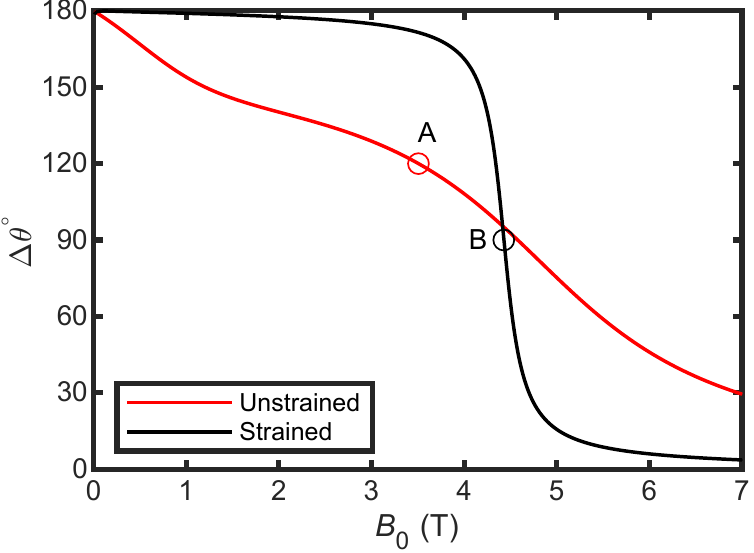}
\caption{Change in nuclear spin quantization direction. $\Delta\theta$ values are plotted as a function of the \vect{B_0} field with the vector oriented $54.7^{\circ}$ away from the SiV axis.}
\label{quant_axes_fixedB0}
\end{figure}

\section{The \siv Hamiltonian}
\label{sec_hamiltonian}
\subsection{Interaction Terms}

We begin by defining the optical ground state of \siv{} Hamiltonian in the lab frame. The combined Hilbert space is a product of the ground state orbitals, and the electron and nuclear spin basis states. The total Hamiltonian is expressed in \cref{eq_total_ham}, and the relevant interaction terms for the electron ($\textbf{H}^{\rm e}$) \& nuclear ($\textbf{H}^{\rm n}$) Zeeman, hyperfine coupling ($\textbf{H}^{\rm hf}$), spin-orbit coupling ($\textbf{H}^{\rm SO}$) and strain ($\textbf{H}^{\rm str}$ defined in the SiV coordinate frame \cite{Hepp2014b}) are expressed in Eqs. (\ref{eq_e_zeeman}) to (\ref{eq_strain}). The \textbf{z} axis is defined along the main symmetry axis of the SiV defect $\langle1 1 1\rangle$. Here, $\boldsymbol{\rm{S}}=[S_{x} ~S_{y} ~S_{z}]^\intercal$ and $\boldsymbol{\rm{I}}=[I_{x}~I_{y}~I_{z}]^\intercal$ are proportional to the Pauli operators for the electron and nuclear spins in the combined Hilbert space. The operators $\boldsymbol{\rm{L}}=[L_{x}~L_{y}~L_{z}]^\intercal$ are the orbital angular momentum operators and are expressed in the spin-orbit basis of $e_{\pm}$; the operators $L_{x}$ \& $L_{y}$ vanish as they only couple the basis states to energy levels much higher in energy \cite{Hepp2014}. The static magnetic field vector is defined as $\boldsymbol{\rm{B_0}}=[B_{x}~B_{y}~B_{z}]^\intercal$. The strain Hamiltonian is also expressed in the spin-orbit basis using the transformation operator $\mathds{T}$ as shown in Eq. (\ref{eq_strain}) \cite{Hepp2014b}. The relevant parameters of the Hamiltonian are summarized in \Cref{ham_constants}. 

\begin{equation}
\label{eq_total_ham}
\textbf{H} = \textbf{H}^{\rm e} + \textbf{H}^{\rm n} + \textbf{H}^{\rm hf} + \textbf{H}^{\rm SO} + \textbf{H}^{\rm str},
\end{equation}
where:
\begin{equation}
\label{eq_e_zeeman}
\textbf{H}^{\rm e}  = \gamma_{e}(\boldsymbol{\rm{B_{0}}}\cdot{}\boldsymbol{\rm{S}}) + \gamma_{L}q L_{z}B_{z}, 
\end{equation}
\begin{equation}
\textbf{H}^{\rm n}  = -\gamma_{n}(\boldsymbol{\rm{B_{0}}}\cdot{}\bf{I}),
\end{equation}
\begin{equation}
\label{eq_hyperfine}
\textbf{H}^{\rm hf}  = A_{||}{S_{z}}{I_{z}} + A_{\perp}({S_{x}}{I_{x}} + {S_{y}}{I_{y}}),
\end{equation}
\begin{equation}
\textbf{H}^{\rm SO} = -\lambda_{\rm SO}L_{z}S_{z},
\end{equation}
\begin{multline}
\label{eq_strain}
\textbf{H}^{\rm str} = 
\mathds{T}
\left(\begin{bmatrix}
\alpha & \beta\\
\beta & -\alpha
\end{bmatrix}\otimes\mathds{1}_{4}\right)
\mathds{T}^{-1},\\
\mathds{T} = \begin{bmatrix}
-1 & -i\\
1 & -i
\end{bmatrix}\otimes\mathds{1}_{4}
\end{multline}

\begin{table}[htp]
\caption{Hamiltonian physical constants for \siv.}
\begin{center}
\begin{tabular}{cll}
\textbf{Parameter} & \textbf{Value} & \textbf{Description}\\
\hline
\hline
$\gamma_e$ & 28 GHz/T & Electron gyromagnetic ratio\\

$\gamma_n$ & -8.46 MHz/T & Nuclear gyromagnetic ratio\\

$A_{||}$ & 70 MHz & Hyperfine  $||$ SiV\textsuperscript{-}-axis \cite{Pingault2017}\\

$A_{\perp{}}$ & 78 MHz & Hyperfine $\perp$ SiV\textsuperscript{-}-axis \cite{Nizovtsev2020,Goss2007}\\

$q$ & 0.1 & Orbital quenching factor\\

$\gamma_{L}$ &  $\gamma_{e}/2$ & Orbital gyromagnetic ratio\\

$\lambda_{\rm SO}$ & 46 GHz & Ground state SO coupling \cite{Rogers2019}\\

$\alpha$ & 150 GHz & $\frac{1}{2}(\epsilon_{xx}-\epsilon_{yy})$ (in-plane \\
 &  & strain anisotropy) \\

$\beta$ & 150 GHz & $\epsilon_{xy}$ (shearing strain)\\

\hline
\end{tabular}
\end{center}
\label{ham_constants}
\end{table}%

In this work we investigate both the unstrained ($\alpha=\beta$=~0 GHz) and strained ($\alpha=\beta$=~150 GHz) regimes of the \siv{} Hamiltonian, as they can lead to different implementations of the spin qubit, i.e., by either driving the spin-orbit separated transition $\omega_1$ or the mixed orbital spin transitions $\omega_2$ [see \cref{b0_optimimum_plot}(a)] \cite{Rogers2014a}. This is a relevant consideration as it is possible to strain engineer \cite{Meesala2018} the silicon vacancy color centers, or fabricate them with naturally unstrained (or minimally strained) \cite{Hepp2014} properties. The values of strain were chosen such that there is sufficient mixing between orbitals, allowing electron spin transitions between spin states. Such values of strain allow the electron spin to undergo a full $\pi$-rotation for an optimised \vect{B_0} field, which need not be re-optimised for higher strain values. The relevant electron spin transitions of interest are illustrated in \Cref{b0_optimimum_plot}(a) for the unstrained and strained Hamiltonians, with transition labeled $\rm\omega_{1}$ and $\rm\omega_{2}$, respectively. These transitions represent electron $\pi$-rotations between eigenstates of the Hamiltonian. Lastly, the contribution of the Jahn-Teller interaction strength is not considered due a relatively negligible strength compared to the $\lambda_{SO}$ and strain \cite{Hepp2014}.

\subsection{Spin Quantization Axis}

The quantization axis of the electron spin depends primarily on the net magnetic field vector \vect{B_{net}}\cite{slichter} -- the sum of the applied field \vect{B_0}, the field induced by spin-orbit coupling \vect{B_{\rm SO}}, and the hyperfine field via coupling to the \si{} nucleus. Thus, with $|$\vect{B_0}$|=0$ T or with \vect{B_{0}}~$\parallel$~[1 1 1], the effective quantization field for the electron and nucleus remains parallel to the defect axis. However, when misalignment of the \vect{B_0} field is introduced, the effective quantization field is no longer parallel to \vect{B_0} due to the competition between the \vect{B_0} and \vect{B_{\rm SO}} vectors \cite{Hepp2014,Rogers2014a}. The quantization field of the electron with respect to the defect symmetry axis is thus expected to vary as a function of the applied \vect{B_0} field. In comparison, the electron is not significantly affected by the magnetic moment of the nucleus as it results in a negligible hyperfine field of $\sim3$ mT, as felt by the electron. 

On the other hand, the electron spin orientation has a significant effect on the nuclear spin orientation due to the electron's larger magnetic moment, resulting in a hyperfine field of $|$\vect{B_{\rm hf}}$|\approx$~4.14~T, as felt by the $^{29}$Si nucleus. Thus, any change in the orientation of the electron spin also significantly affects the quantization field of the nuclear spin. Moreover, as the spin-orbit coupling is a relativistic effect experienced by the orbiting electron, the nuclear spin only experiences the applied \vect{B_0} field and the hyperfine field \vect{B_{\rm hf}}. In summary, the main effect is that the nuclear spin eigenstate orientations are no longer strictly parallel to each other when the electron spin is flipped. This is the origin of the anisotropy-like effect of the nuclear spin quantization fields, captured by the angle $\Delta\theta$, which enables indirect control to become possible even in the absence of explicit hyperfine anisotropy. The relevant static field vectors of interest are illustrated in \Cref{b0_optimimum_plot}(b).

We investigate the change in orientation of the quantization fields of the nuclear spin represented by $\Delta\theta$ as a result of the relevant electron spin flipping transitions $\ket{\uparrow}\leftrightarrow \ket{\downarrow}$, illustrated in \Cref{b0_optimimum_plot}(a). We plot $\Delta\theta$ as a function of the \vect{B_0} field intensity and orientation for these transitions in \Cref{b0_optimimum_plot}(c) \& (d), for the unstrained and strained Hamiltonians, respectively. Closer inspection of these plots shows that there are ideal parameters for the \vect{B_0} field that result in $\Delta\theta$ close to $90^{\circ}$. This is favourable as the efficiency of the indirect control technique is optimised by maximising this angle \cite{Khaneja2007}. 

As an example we illustrate the exact spin eigenstates of the Hamiltonian, showing the quantization axes orientations in \Cref{quant_axes}, with respect to the \siv{} main symmetry axis. These orientations are plotted with respect to the defect axis, for both cases of strain and with the \vect{B_0} field orientated at $54.7^{\circ{}}$ (corresponding to [1 0 0] vector orientation). Lastly, the $\Delta\theta$ values for the nuclear spin corresponding to the relevant electron transitions [illustrated in \Cref{b0_optimimum_plot}(a)] are plotted in \Cref{quant_axes_fixedB0} as a function of $|$\vect{B_0}$|$ applied at an angle $54^{\circ{}}$ away from the \siv{} symmetry axis.

\begin{figure*}[htbp]
\includegraphics{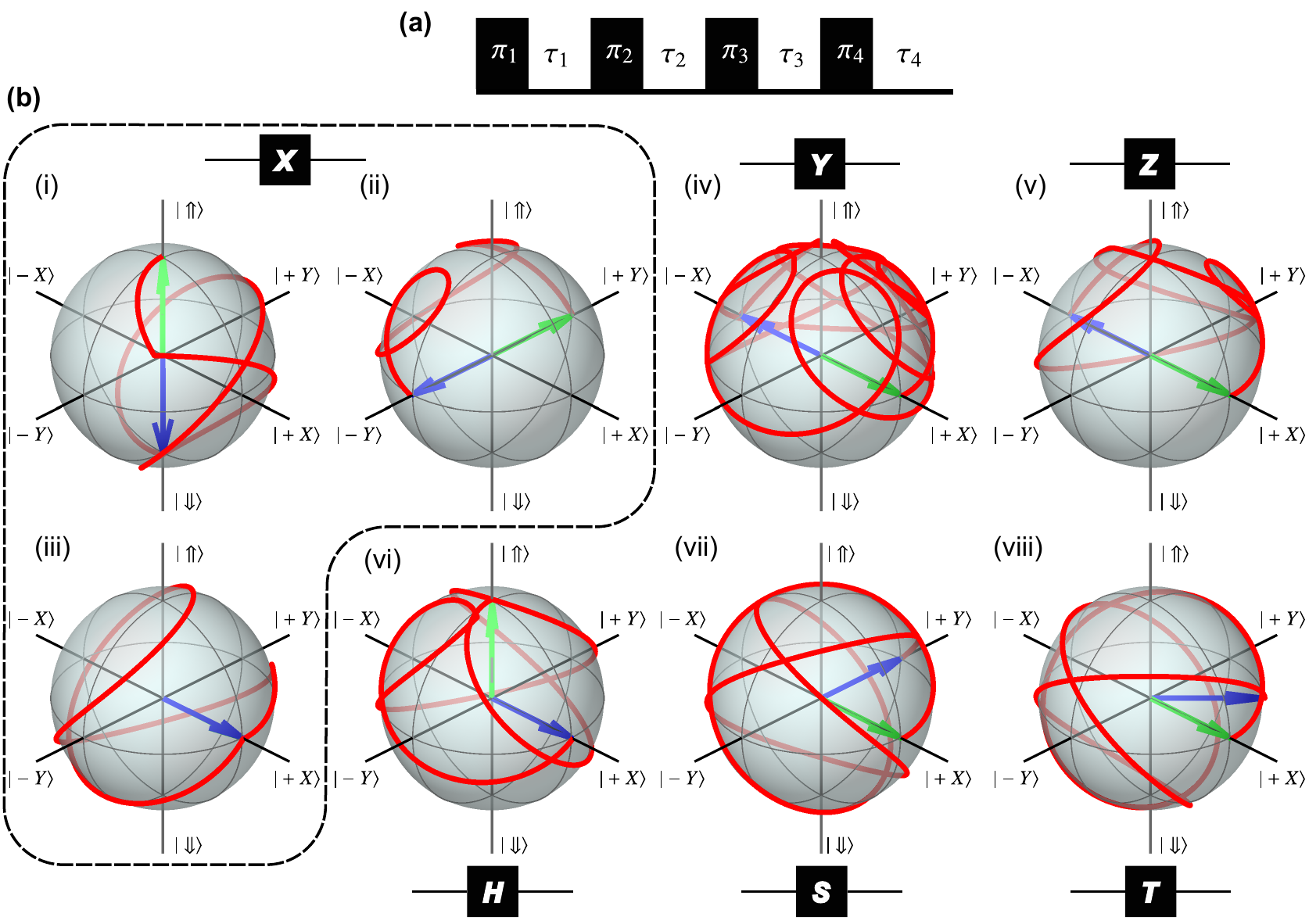}
\caption{Indirect control (IC) of the nuclear spin. 
(a) Series of electron $\pi$-rotations separated by delays $\tau$ constituting a typical IC-gate sequence. 
(b) (i-iii) Bloch sphere representations of the nuclear spin trajectories, showing the action of the \textit{\textbf{X}} gate on the nuclear spin initialised in the $\ket{\Uparrow}$, $\ket{+X}$ \&  $\ket{+Y}$ states, respectively. The time evolved trajectories of the spin in the rotating frame are plotted in solid red on the Bloch sphere. Green arrows represent the initial state of the nuclear spin before the gate sequence is applied, whereas blue arrows represent the final state. (iv-viii) Demonstrations of \textbf{\textit{Y}}, \textbf{\textit{Z}}, Hadamard, \textbf{\textit{S}}-phase and \textbf{\textit{T}}-phase gates.}
\label{bloch_trajectory_all}
\end{figure*}

\section{Indirect Control}
\label{sec_indirect}
For a proof-of-principle demonstration we assume an unstrained SiV center in a [1 0 0] diamond sample, resulting in a \vect{B_0} angle of $54.7^{\circ}$ relative to the defect symmetry axis. Informed by \Cref{quant_axes_fixedB0}, we choose $|$\vect{B_0}$|=3.5$~T such that the change in quantization field orientation $\Delta\theta$ is $120^{\circ{}}$, which is equivalent to $\Delta\theta=60^{\circ{}}$. Such an angle allows for complete control of the nuclear spin on the Bloch-sphere with just four electron $\pi$-rotations \cite{Khaneja2007}, enabling us to construct the required IC gate. The advantage of limiting IC gates to using electron $\pi$-rotations is twofold: (i) the gate construction and optimisation is less computationally intensive and (ii) the electron spin is mostly in an eigenstate and is thus not greatly subject to dephasing, which would ultimately propagate to the nuclear spin through the hyperfine coupling. Although there are methods to implement self-protected IC gates using dynamical decoupling techniques in the presence of noise \cite{Liu2013}, this would introduce additional complexity which is not required for our demonstration.

A sequence of electron $\pi$-rotations and free precession delays ($\tau_{1-4}$) [illustrated in \Cref{bloch_trajectory_all}(a)] is used to actuate on the nuclear spin, evolving the system towards a deterministic target spin state. The gate sequence optimisation is performed with Matlab's `Surrogate Optimisation' solver, using an optimisation routine that considers the overlap of the system's final time-evolved optimised state $|\psi_{\rm opt}\rangle$ and the target state $|\psi_{\rm T}\rangle$ given as $|\langle\psi_{\rm opt}|\psi_{\rm T}\rangle|^{2}$, along with the total gate time as the optimisation goal. We track the time evolution of the nuclear spin in its rotating reference frame, in order to properly derive the correct phase manipulation of the qubit gates.

\subsection{Single Qubit Gates}
\label{qubit_gates}

Two-axes control of a qubit requires control of both the zenith and azimuthal angles over the Bloch sphere. Conventionally, phase control of a spin qubit is performed using IQ modulation of the MW or RF pulses \cite{Sewani2020}-- which is not possible in the absence of such oscillating fields. Importantly, simple precession of the nuclear spin around the primary quantization axis merely results in a global phase accumulation which is not relevant for quantum information processing. Fortunately, as we show here, phase control of the nuclear spin does not impose any further requirements on the electron spin control apart from simple $\pi$-rotations. In this way, the phase is accumulated as the spin precesses around the secondary quantization axis which detunes the spin from the resonance frequency.

We consider the application of a standard set of single qubit gates on the nuclear spin which include phase gates, implemented in simulation as instantaneous electron $\pi$-rotations. The nuclear spin IC gates are numerically optimised by enforcing that two seperate qubits each initialised in a different state correctly evolve to their respective final states in their rotating frames. Examples of gate actions on a pair of initial states are $\ket{+X}\leftrightarrow\ket{+X}$ and $\ket{\Uparrow}\leftrightarrow\ket{\Downarrow}$ for the \textit{\textbf{X}} gate, where $\ket{\pm{}X} = (\ket{\Uparrow} \pm \ket{\Downarrow})/\sqrt{2}$.

Further, it is enforced that gate operations may only be performed when the lab frame is in phase with the rotating frame. This is to account for the fact that the quantization fields remain stationary in the lab frame. To achieve this, an appropriate delay is padded to the end of the optimised gate sequence, such that the rotating frame returns to phase with the lab frame at the end of the sequence. This padding delay is given as $\tau_{4} = \frac{\ceil*{T_{\rm gate}f_{\rm RF}}}{f_{\rm RF}} - T_{\rm gate}$, where $f_{\rm RF}$ is the nuclear spin precession frequency, $T_{\rm gate}$ is the total optimised gate time and the $\ceil*{}$ brackets represent a ceiling function. This ensures that there is a unique optimised gate sequence for a given gate operation.

We demonstrate the following gates: \textit{\textbf{X}, \textbf{Y}, \textbf{Z}, \textbf{H}, \textbf{S}} \& \textit{\textbf{T}}, with the gate durations summarised in \Cref{table_gate_times} for the unstrained (`A') and strained (`B') Hamiltonian parameters, also indicated in \Cref{quant_axes_fixedB0}. The gates are simulated with high fidelity (\textit{F}~$\ge 0.98$), with diminishing returns expected for longer optimisation routines. 

We find the total gate times to be at the order of hundreds of nanoseconds, significantly faster than those that can be achieved with NMR techniques. For example, the length of the IC \textit{\textbf{Y}} gate for the unstrained Hamiltonian (`A') is equivalent to that achieved via an NMR Rabi frequency of $\sim1.6$ MHz. Such high speed control would typically require an applied radio-frequency $B_1$ field of $\sim 400$~mT, which is impossible to achieve in millikelvin cryostats. Note that the gate times are generally higher for the strained case despite $\Delta\theta=90^{\circ{}}$. This is because the gate durations are limited by lower precession frequencies of the nuclear spin as discussed in \Cref{nuclear_precession_times}. 

Lastly, we can find the inverse gates \textit{\textbf{S\textsuperscript{-1}}} \& \textit{\textbf{T\textsuperscript{-1}}} for uncomputation operations \cite{Nielson2011} using the same method described in \cref{qubit_gates}. 

\begin{table}[htbp]
\caption{Simulated single qubit IC gate parameters and durations for the nuclear spin in the unstrained and strained Hamiltonians corresponding to labels A \& B, respectively, indicated in \Cref{quant_axes_fixedB0} \& \Cref{b0_optimimum_plot}(c),(d). The average simulated gate fidelity \textit{\textbf{F}} is over 0.98.}
\label{table_gate_times}
\resizebox{\columnwidth}{!}{%
\begin{tabular}{c|c|r r r r|r|c}
    \hline
    \hline
    \textbf{Strain} & \multirow{2}{*}{\textbf{Gate}} & \multirow{2}{*}{$\tau_{1}$~(ns)} &  \multirow{2}{*}{$\tau_{2}~(ns)$}  & \multirow{2}{*}{$\tau_{3}~(ns)$} & \multirow{2}{*}{$\tau_{4}~(ns)$} & \textbf{Total} & \multirow{2}{*}{\textit{\textbf{F}}} \\
     \textbf{Case} &  &  &    &  &  &(ns) & \\
    \hline 
    \multirow{9}{*}{\rotatebox{90}{Unstrained (A)~}} & \textit{\textbf{X}} & 2.99   & 172.96  & 22.66 & 10.46 & 209.06 & $\ge$0.99\\
    & \textit{\textbf{Y}} & 11.73  & 197.03  & 84.09 & 20.73 & 313.59 & $\ge$0.99\\
    & \textit{\textbf{Z}} & 21.34  & 96.27   & 21.29 & 17.89 & 156.79 & $\ge$0.99\\
    & \textit{\textbf{H}} & 29.33  & 111.23  & 10.55 & 5.69 & 156.79 & $\ge$0.99\\
    & \textit{\textbf{S}} & 5.18   & 90.85   & 35.24 & 25.52 & 156.79 &
    $\ge$0.99\\
      & \textit{\textbf{S\textsuperscript{-1}}} & 31.73   & 238.11  & 1.62 & 42.13  & 313.59 & $\ge$0.99\\
    & \textit{\textbf{T}} & 4.49   & 88.29   & 34.56 & 29.47 & 156.79 & $\ge$0.99\\
    & \textit{\textbf{T\textsuperscript{-1}}} & 90.26  & 184.23  & 60.14 & 31.22 & 365.86 & $\ge$0.99\\
 \hline
    \multirow{9}{*}{\rotatebox{90}{Strained (B)~}} & \textit{\textbf{X}} & 245.39 & 1.49 & 401.84 & 647.73 & 1296.50 & $\ge$0.99\\
    & \textit{\textbf{Y}} & 119.27 & 329.26 & 192.64 & 7.06 & 648.23 & $\ge$0.98\\
    & \textit{\textbf{Z}} & 86.42  & 1.77   & 233.88 & 326.16 & 648.23& $\ge$0.99\\
    & \textit{\textbf{H}} & 28.76  & 162.03 & 196.84 & 260.61 & 648.23& $\ge$0.98 \\
    & \textit{\textbf{S}} & 128.51 & 1.05 & 31.65 & 487.02 & 648.23 & $\ge$0.99\\
    & \textit{\textbf{S\textsuperscript{-1}}} & 60.16   & 195.47  & 33.49 & 358.1  & 648.23 & $\ge$0.99\\
    & \textit{\textbf{T}} & 79.09  & 1.63 & 1.00  & 566.50  & 648.23 & $\ge$0.98\\
    & \textit{\textbf{T\textsuperscript{-1}}} & 60.14  & 223  & 60.14 & 31.22 & 648.23 & $\ge$0.99\\
    \hline
    \hline
\end{tabular}
}
\end{table}

\begin{figure}[htb]
\includegraphics[keepaspectratio]{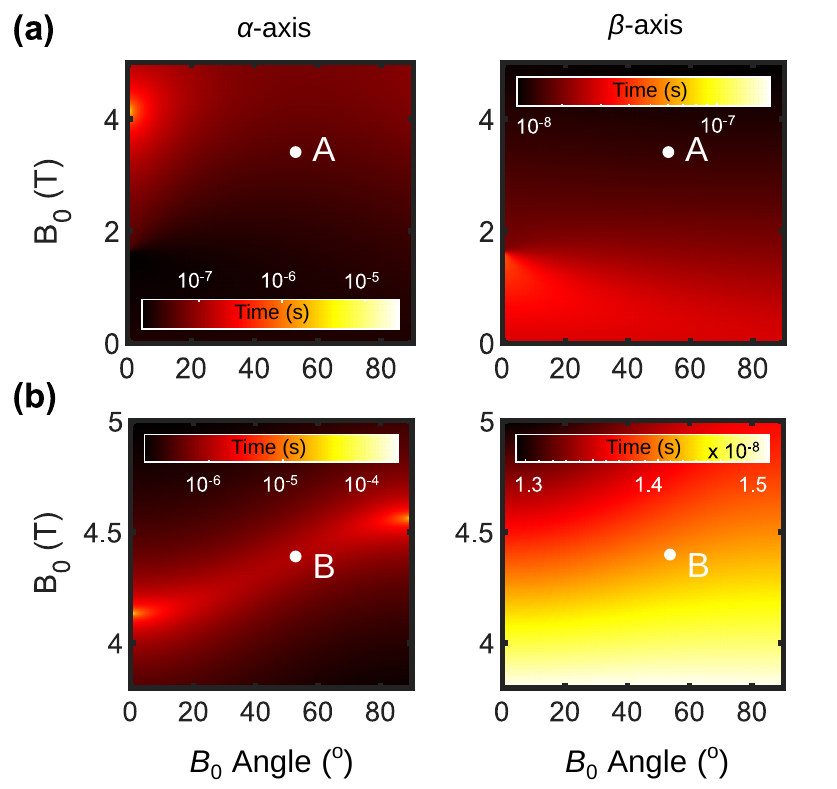}
\caption{Nuclear spin precession. 
(a) Nuclear spin precession periods of the primary $\alpha$ and secondary $\beta$ quantization axes plotted as a function of the \vect{B_0} field for the unstrained case. 
(b) Same plots as (a) but for the strained case. These plots are to be used as reference together with \Cref{b0_optimimum_plot}(c) \& (d).}
\label{precession_times}
\end{figure}

\section{Discussion}
\label{discussion}

\subsection{High-speed electron $\pi$-rotations}
In this work we have shown simulations wherein instantaneous electron $\pi$-rotations are used to actuate on the nuclear spin orientation. Fast electron switching is required to minimally disturb both the electron and nuclear spins during the switching period, and to keep the electron switching fast compared to the nuclear spin timescales. Practical realisations of such high-speed electron Rabi frequencies involve high microwave frequency $B_1$ fields, which though achievable, would require high-Q factor resonators with contradictorily long ring-down times (order of 100 ns). This limits the suitability of magnetic resonance techniques for coherent electron spin control. Alternatively, high electron Rabi frequencies can be achieved using coherent acoustic control via surface acoustic waves \cite{Maity2020}, or using coherent all-optical control of the electron spin state \cite{Becker2018b,Takou2021}. 

All-optical control is particularly favourable, as it does not involve the need for fabrication of an on-chip antenna or acoustic transducer. This would allow for precise, targeted nuclear spin qubit control, taking advantage of the state-of-the-art optical confocal microscopy methods \cite{kuhlmann}. The use of nanowatt scale optical power renders this method readily compatible with ultra low temperature cryostats such as dilution refrigerators with a cooling power of 10-500~$\mu{}$W \cite{bluefors}. Lastly, optical manipulation of the electron spin offers the potential of picosecond timescale $\pi$-rotations \cite{Berezovsky2008}, which justifies the instantaneous spin rotations employed in our simulations.

\subsection{Nuclear gate timescale}

\label{nuclear_precession_times}
The efficiency of indirect control directly depends on the delay times between switching of the quantization axes, which are determined by the precession frequency of the nuclear spin under the influence of its quantization field. A \vect{B_0} field in a non-optimal orientation may negate the benefits of indirect control. We plot the nuclear spin precession period as a function of magnetic field orientation, for the primary and secondary quantization axes as well as for the unstrained and strained cases in \Cref{precession_times}. These plots are to be used as reference for optimising the \vect{B_0} field parameters while also considering the resulting change in quantization axes as shown in \Cref{b0_optimimum_plot}(c)\&(d). For the unstrained \siv{} the precession times can range up to few microseconds for both the primary and secondary quantization fields [cf. \Cref{precession_times}(a)]. Whereas, for the strained \siv{} the precession period is increased up to tens of microseconds with a \vect{B_0} field aligned either parallel or perpendicular to the \siv{} defect axis [cf. \Cref{precession_times}(b)]. For both the strain and unstrained cases, it is comparatively favourable to align the \vect{B_0} field at $54.7^{\circ{}}$ as used in this work, given that an appropriate $|$\vect{B_0}$|$ is chosen.

\section{Conclusion}
 We have proposed a path towards fast and coherent control of the host nuclear spin in the \siv{} defect in diamond through indirect control methods using simulations of time-optimised single qubit gates. We have shown that the large spin-orbit interaction in the \siv{} Hamiltonian replaces the need for an appreciable hyperfine anisotropy required for indirect control shown in literature. We demonstrate various simulated single qubit IC gate times for both the unstrained and strained Hamiltonians as proof of principle, considering that both strain cases of the silicon vacancy centers can be found naturally or through strain engineering. By using instantaneous electron $\pi$-rotation sequences we demonstrate one-qubit nuclear spin gates at megahertz rates. We envision the exciting possibility of all-optical nuclear spin control in SiV defects in diamond.  Finally, we expect indirect control techniques to be carried over to other spin systems that exhibit similar electronic properties such as the negatively-charged \textsuperscript{73}GeV centers in diamond \cite{Siyushev2017,vladimir2016}.
 
\section*{Acknowledgements}
We thank V.K. Sewani for useful discussions. We acknowledge support from the Australian Research Council (CE170100012, FL190100167). H.H.V. acknowledges support from the Sydney Quantum Academy. C.A. and A.L. acknowledge support from the University of New South Wales Scientia program.

\bibliography{references}
\end{document}